\documentclass[showpacs,aps,prl,reprint,superscriptaddress,preprintnumbers,nolinenumbers]{revtex4-1}

\usepackage{amsmath}
\usepackage{amssymb}
\usepackage{graphicx}
\usepackage{dcolumn}
\usepackage[dvipsnames]{xcolor}
\usepackage{amsthm,amssymb}
\usepackage{mathrsfs}
\usepackage{color}
\usepackage{endnotes}
\usepackage{bm}
\usepackage{epsfig}
\usepackage{array}
\usepackage{ulem}
\usepackage{soul}

\newcommand{\etal}{{\it et al.}}
\newcommand{\ie}{{\it i.e.\ }}

\newcommand{\parallelsum}{\mathbin{\!/\mkern-5mu/\!}}

\begin{document}

\title{Current direction dependent magnetotransport in CuTe}

\author{Ying~Kit~Tsui}
\affiliation{Department of Physics, The Chinese University of Hong Kong, Shatin, Hong Kong, China}
\author{C. N.~Kuo}
\affiliation{Department of Physics, National Cheng Kung University, Tainan 70101, Taiwan}
\author{C.~E.~Hsu}
\affiliation{Department of Physics, Tamkang University, Tamsui, New Taipei, 25137, Taiwan}
\author{Wei~Zhang}
\author{Wenyan~Wang}
\affiliation{Department of Physics, The Chinese University of Hong Kong, Shatin, Hong Kong, China}
\author{Shanmin~Wang}
\affiliation{Department of Physics, Southern University of Science and Technology, Shenzhen, Guangdong 518055, China}
\author{Wing~Chi~Yu}
\affiliation{Department of Physics, City University of Hong Kong, Kowloon, Hong Kong, China}
\author{H.~C.~Hsueh}
\affiliation{Department of Physics, Tamkang University, Tamsui, New Taipei, 25137, Taiwan}
\author{C. S.~Lue}
\affiliation{Department of Physics, National Cheng Kung University, Tainan 70101, Taiwan}
\author{Swee~K.~Goh}
\email[]{skgoh@cuhk.edu.hk}
\affiliation{Department of Physics, The Chinese University of Hong Kong, Shatin, Hong Kong, China}

\date{\today}

\begin{abstract}
Despite being a layered, easily-exfoliated compound, copper monotelluride (CuTe) features an unusual quasi-one-dimensional charge density wave below $T_{\rm CDW}\approx335$~K. Within a CuTe layer, the electrical resistivity depends sensitively on the direction of the electrical current. Here, we use magnetotransport to probe the metallic state of CuTe with two distinct in-plane current directions. When the current flows along the $a$-axis ($I\parallelsum a$), the magnetoresistance exhibits a downward curvature as the magnetic field increases. On the other hand, when the current is along the $b$-axis ($I\parallelsum b$), the magnetoresistance shows the opposite curvature. Our analysis uncovers a violation of Kohler scaling, but only for $I\parallelsum a$. Shubnikov-de Haas oscillations are detected at low temperatures. Our results shed light on the nature of the metallic state in CuTe with the development of the charge density wave.

\end{abstract}

\maketitle

\section{I.~INTRODUCTION}
Charge density wave (CDW) is an electronic state that involves a periodic modulation of conduction electron density in real space \cite{Gruner2018Book}. Recently, CDW has been attracting attention partly because of its observation in the proximity of superconducting state \cite{Chang2012YBCO, Dreher2021NbSe2, Kusmartseva2009TiSe2, Kudo2010SrPt2As2, Li2017TaS2, Chen2021CsV3Sb5, Neupert2022AV3Sb5, Wang2021CsV3Sb5, Sipos2008TaSe2, Liu2016TaSe2}. Understanding the interplay between the CDW phase and the superconductivity can shed light on the superconducting state, potentially enabling a fine-tuning of the superconducting properties. For example, the recently discovered kagome superconductors AV$_3$Sb$_5$ (A = K, Rb, and Cs) are found within the CDW phase, and the suppression of the CDW state drastically enhance the superconducting transition temperature ($T_c$) of AV$_3$Sb$_5$ \cite{Chen2021CsV3Sb5, Neupert2022AV3Sb5, Wang2021CsV3Sb5}. On the other hand, in 1T-TaS$_2$, the suppression of the CDW phase does not enhance $T_c$ significantly \cite{Sipos2008TaSe2, Liu2016TaSe2}. Thus, superconductivity appears to be competing with the CDW state in AV$_3$Sb$_5$ but not in 1T-TaS$_2$. To gain further insights, understanding how the CDW emerges is crucial.

Copper monotelluride (CuTe) has recently been shown to host a rare quasi-one-dimensional (quasi-1D) CDW phase below $T_{\rm CDW}\approx335$~K \cite{Stolze2013CuTe, Zhang2018CuTe, Kuo2020CuTe, Wang2021CuTe, Wang2022CuTe, Li2022CuTe, Kim2019CuTe, Cudazzo2021CuTe}. Above $T_{\rm CDW}$, CuTe adopts an orthorhombic structure consists of weakly bonded layers along the $c$ axis \cite{Pertlik2001CuTe, Seong1994CuTe, Stolze2013CuTe}. Interestingly, the formation of the CDW state only involves the distortion of Te-atom chains along the $a$ axis, in which the Te atoms form groups of two or three atoms \cite{Stolze2013CuTe, Zhang2018CuTe, Kim2019CuTe, Cudazzo2021CuTe, Li2022CuTe, Wang2022CuTe}. By applying pressure, the CDW state can be suppressed, and superconductivity emerges~\cite{Wang2021CuTe, Wang2022CuTeb}. Therefore, CuTe is another platform to explore the interplay between the CDW state and superconductivity. When CuTe is cooled across $T_{\rm CDW}$, Seebeck coefficient exhibits a clear kink and decreases with a faster rate \cite{Kuo2020CuTe}. Similarly, an anomaly is detected in thermal conductivity at $T_{\rm CDW}$ \cite{Kuo2020CuTe}. These behaviours are consistent with  partially gapped Fermi surfaces because of the CDW formation~\cite{Kim2019CuTe}. Furthermore, the specific heat jump is larger than the mean-field expectation at the transition temperature, supporting the possibility of a strong-coupling CDW~\cite{Kuo2020CuTe}. Angle-resolved photoemission spectroscopy (ARPES) measurement, together with calculated electron scattering susceptibility and phonon dispersion, revealed the Fermi surface nesting with a single $\vec{q}$ vector~\cite{Zhang2018CuTe}. Besides, the observed phonon-mode shift in Raman scattering when approaching the CDW transition temperature points to strong electron–phonon coupling in the formation of the CDW~\cite{Wang2022CuTe}. Thus, both the electron-electron scattering associated with Fermi surface nesting and electron-phonon coupling are likely to be active in the formation of the CDW phase in CuTe.

ARPES data show that the quasi-1D Fermi sheets parallel to the $k_y$ direction are gapped out in the CDW state, while the quasi-2D pockets are less affected \cite{Zhang2018CuTe}. Thus, the electronic transport is expected to be anisotropic. Furthermore, if the setting in of the CDW phase introduces new periodicity, the quasi-2D Fermi pockets can be reconstructed because of the new Brillouin zone associated with the CDW phase. To explore this scenario, magnetic quantum oscillations can be conducted and compared with the calculated non-CDW Fermi surfaces. Quantum oscillation frequencies gauge the size of Fermi surfaces. Hence, if  Fermi surfaces are reconstructed, the associated quantum oscillation frequencies will be smaller than the frequencies extracted from the calculated non-CDW Fermi surfaces.

To explore the anisotropic transport and to detect quantum oscillations in CuTe, we prepared  devices to conduct magnetotransport measurements on high-purity single-crystalline thin flakes of CuTe. We detected obvious difference when the in-plane current directions are varied. Furthermore, we succeeded in detecting 
clear magnetic quantum oscillations in resistivity (aka the Shubnikov-de Haas (SdH) effect). To the best of our knowledge, this is the first detection of the SdH effect in CuTe, offering a chance to investigate the fermiology of CuTe in the CDW state. This can also provide clues for understanding the magnetoresistance data of CuTe.

\section{II.~METHODS}
CuTe single crystals were grown using Te self-flux method as reported earlier~\cite{Kuo2020CuTe}. Several single crystals were ground to obtain a powder specimen for synchrotron X-ray powder diffraction (SXRD) \cite{Footnote1}. For transport measurements, cleaved samples were transferred onto silicon-based substrates pre-patterned with gold electrodes. To prepare these electrodes, a small piece of silicon wafer with a 300-nm-thick oxide layer was used. A photomask was then used to expose a Hall bar pattern for photolithography. After gold deposition, the silicon substrate with gold Hall bar pattern was ready for sample transfer. The substrates provided physical support and electrical contacts that facilitated resistivity measurements. The low temperature and high magnetic field were achieved using a Physical Property Measurement System (PPMS) by Quantum Design. A single-axis rotator was installed to vary the orientation of the samples with respect to the magnetic field direction. Resistivity measurements were performed with the resistivity option of the PPMS except the quantum oscillation, for which a Stanford Research 830 lock-in amplifier was used. The first-principles electronic structures of quasi-one dimensional CuTe were calculated using the Quantum Espresso code \cite{Giannozzi2020QuantumExpresso} with the ultrasoft pseudopotentials in the framework of density functional theory (DFT). FermiSurfer was used to display the calculated Fermi surface \cite{Kawamura2019FermiSurfer}. Quantum oscillation frequencies were extracted using Supercell K-space Extremal Area Finder (SKEAF)~\cite{Rourke2012SKEAF}. Further computational details are presented in the Supplemental Material \cite{SUPP}. 

\section{III.~RESULTS AND DISCUSSIONS}
\begin{figure}[!t]\centering
      \resizebox{9cm}{!}{
              \includegraphics{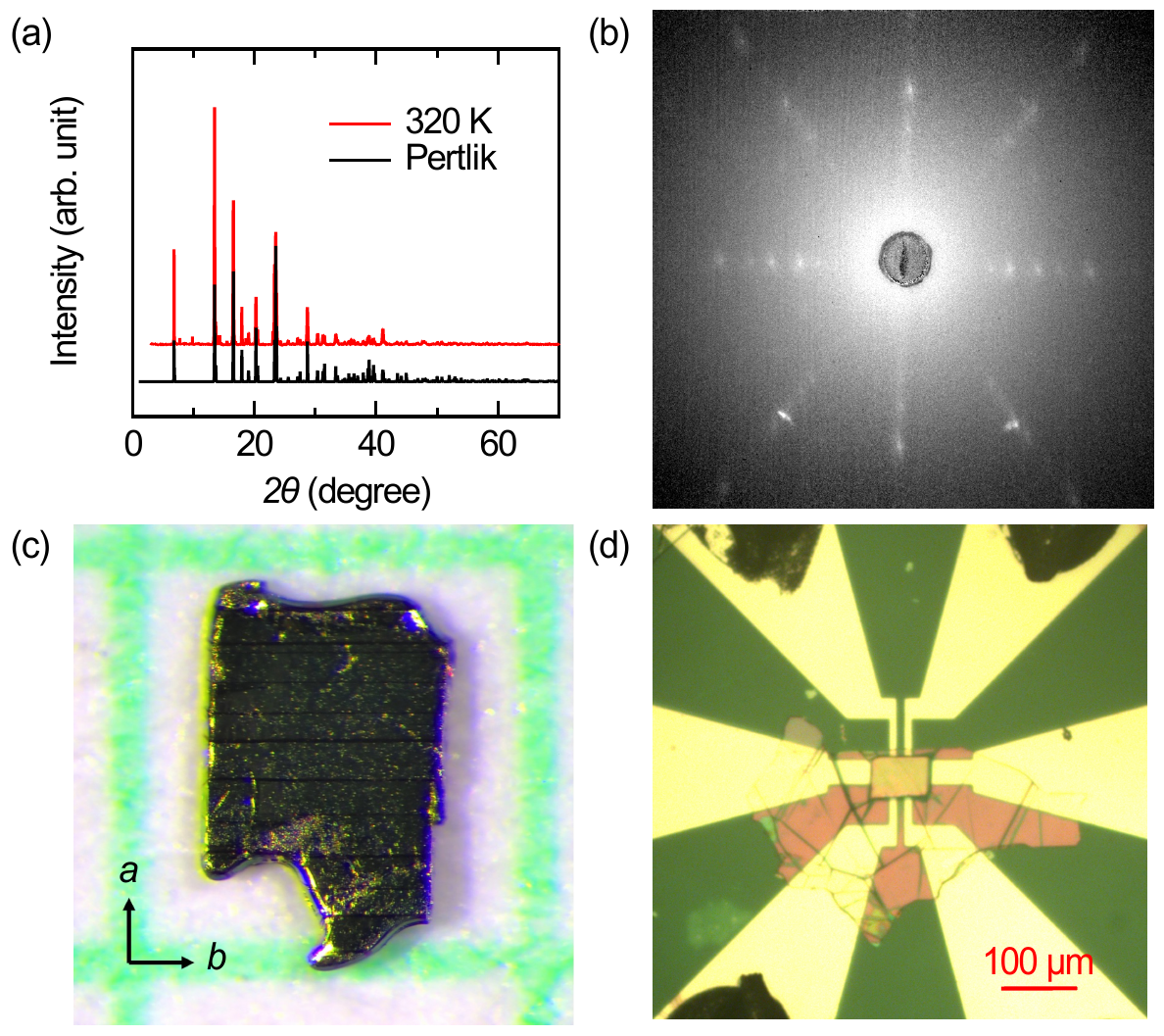}}		
              \caption{\label{fig1}  
(a) Powder X-ray diffraction spectrum of CuTe in comparison with the data from Pertlik \cite{Pertlik2001CuTe}. (b) Laue diffraction image captured along the [001] direction. (c) Typical CuTe crystal. The arrows indicate the direction of $a$ and $b$ lattice vectors. (d) A thin flake on the patterned substrate. The sample is at the centre of the photograph, and the shinier areas are the gold electrodes. The electrical current is flowing from the left electrode to the right electrode. A layer of h-BN was added on top.}
\end{figure}

Figure~\ref{fig1}(a) displays the powder X-ray diffraction spectrum of CuTe. Clear peaks can be seen, and they can be indexed to $Pmmn$ space group, confirming the structure of the sample \cite{Stevels1971CuTe, Seong1994CuTe, Pertlik2001CuTe}. In Fig.~\ref{fig1}(b), the Laue diffraction pattern exhibits well-defined spots, demonstrating the high crystallinity of the single crystals used for this study. Figure~\ref{fig1}(c) is a photograph of a typical single crystal placed on a paper with 1 mm $\times$ 1 mm grids. The crystal is position with the $ab$ plane parallel to the paper. Even at the visual level, stripes along the $b$ axis can be seen. CuTe crystals can be mechanically exfoliated. The resultant thin flakes provide the advantage of enhanced resistive signals due to the optimal geometric factor. Figure~\ref{fig1}(d) shows a single crystal of CuTe positioned on the pre-patterned conductive electrodes. A layer of h-BN was added to protect the thin flake from the atmosphere. As will be discussed below, such a device provides superior resistive signals, enabling the observation of the SdH oscillations.
\begin{figure}[!t]\centering
      \resizebox{9cm}{!}{
 \includegraphics{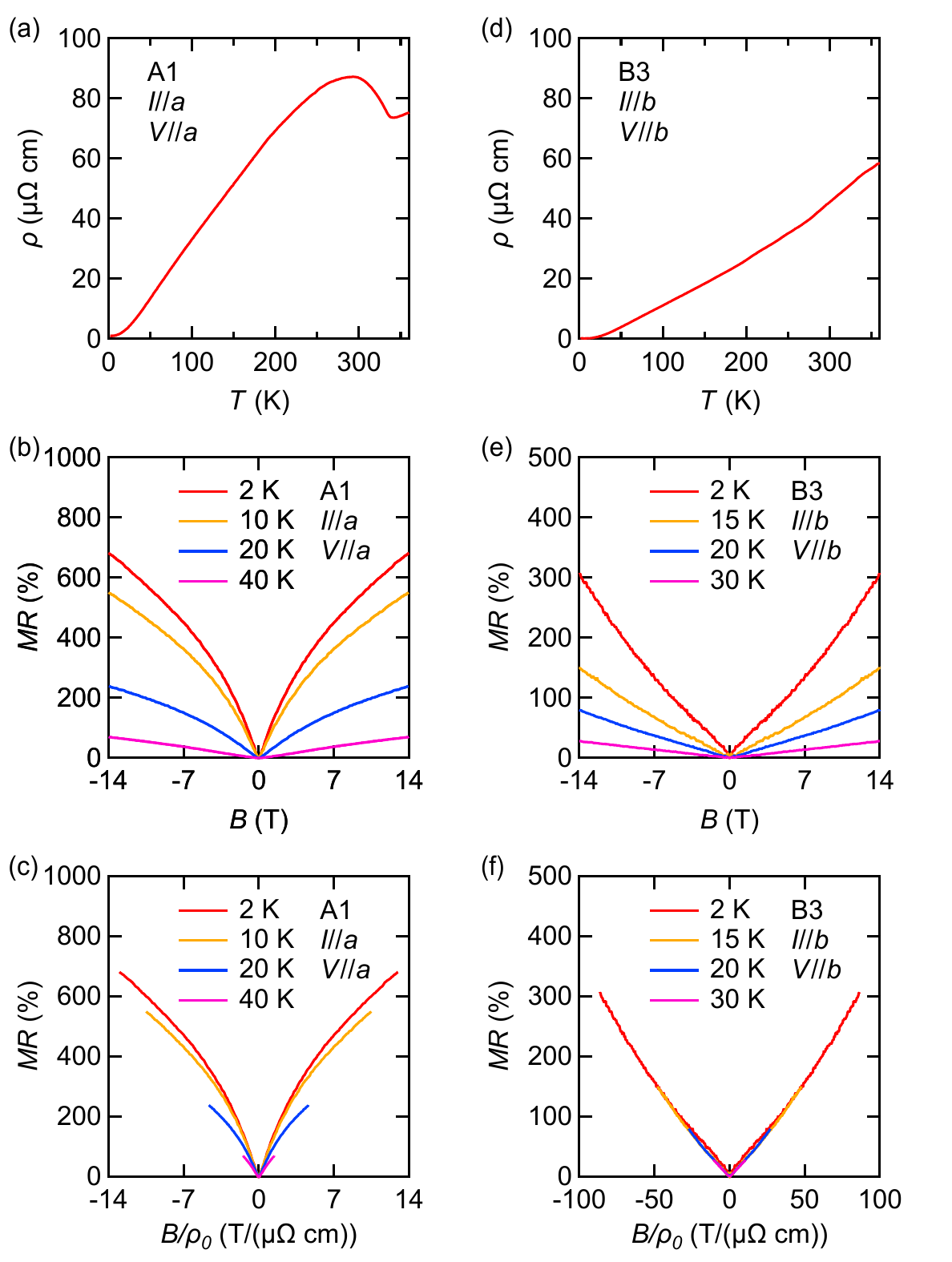}}       
 \caption{\label{fig2} 
(a) Temperature dependence of the electrical resistivity of sample A1. (b) Magnetoresistance along the $a$ axis at different temperatures. (c) Kohler scaling plot for (b). (d) Temperature dependence of the electrical resistivity of sample B3. (e) Magnetoresistance along the $b$ axis at different temperatures. (f) Kohler scaling plot for (e).}
\end{figure}

The morphology of the crystals allows easy identification of the crystalline axes, facilitating the preparation of experiments to investigate the transport properties with current flowing along two orthogonal planar directions. For instance, the thin flake shown in Fig.~\ref{fig1}(d) is arranged so that the current flows along the crystalline $a$ axis, with the longitudinal voltage measured along the same axis. Figure~\ref{fig2}(a) shows the temperature dependence of resistivity of sample A1, which has current flowing along the $a$ direction while Fig.~\ref{fig2}(d) shows the temperature dependence of resistivity of sample B3, which is configured with current along the $b$ direction. Both A1 and B3 exhibit a metallic $\rho(T)$ down to 2~K, the base temperature of our experiment. The residual resistance ratio (RRR), defined as $\rho(300~{\rm K})/\rho(2~{\rm K})$, is about 86 and 277 for A1 and B3, respectively. Such high RRRs demonstrate the usage of high-quality samples in our study. The onset of the CDW transition manifests itself as a pronounced resistive upturn in A1. However, the resistivity is almost smooth across $T_{\rm CDW}$ in B3. This striking dependence on the current direction is in agreement with the data collected by Li \etal~\cite{Li2022CuTe} and Wang \etal~\cite{Wang2022CuTe}. We also confirm the current-direction dependent $\rho(T)$ by Montgomery Method, as presented in the Supplemental Material~\cite{SUPP}. Thus, our device-based approach produces consistent data for the thin flake.

The anisotropy in the temperature dependence of the resistivity motivates us to investigate the magnetoresistance (MR) with different current directions. Figures~\ref{fig2}(b) and 2(e) show the MR of A1 and B3 respectively, measured up to 14~T at different temperatures. At 2~K and 14~T, the magnitude of MR ($=[\rho(14~{\rm T})-\rho_0]/\rho_0 \times 100\%$) reaches 682\% and 305\% for A1 and B3, respectively. Here, $\rho_0$ is the resistivity at the zero field. The large MR is again consistent with the high sample quality. Interestingly, the overall field dependence of the MR depends on the current direction. When the current is along the $a$ axis (sample A1), the MR shows an overall downward curvature and a tendency of saturation at high field. On the contrary, the MR shows an upward curvature in B3. With an increasing temperature, the magnitude of MR decreases in both A1 and B3, but the overall curvature is similar to the datasets at 2~K.

To see the similarity of the curvature in MR($B$) at various temperatures, we construct a Kohler plot. Plotting the magnetoresistance against $B/\rho_0$, the MR curves of B3 at different temperatures can be well represented by a single curve (Fig.~\ref{fig2}(f)), indicating the observation of Kohler scaling. However, Kohler's rule is violated in A1, as evidenced in Fig.~\ref{fig2}(c). The Kohler scaling in B3 suggests the universal scattering mechanism when the current is along $b$. The violation of the Kohler scaling when current is along $a$ suggests the changing of the carrier density or the scattering mechanism as the temperature changes. The changing of the carrier density is plausible because ARPES detected the increasing of the CDW gap size for the quasi-1D Fermi sheets parallel to the $k_y$ axis as the temperature decreases \cite{Zhang2018CuTe}.

\begin{figure}[!t]\centering
       \resizebox{9cm}{!}{
\includegraphics{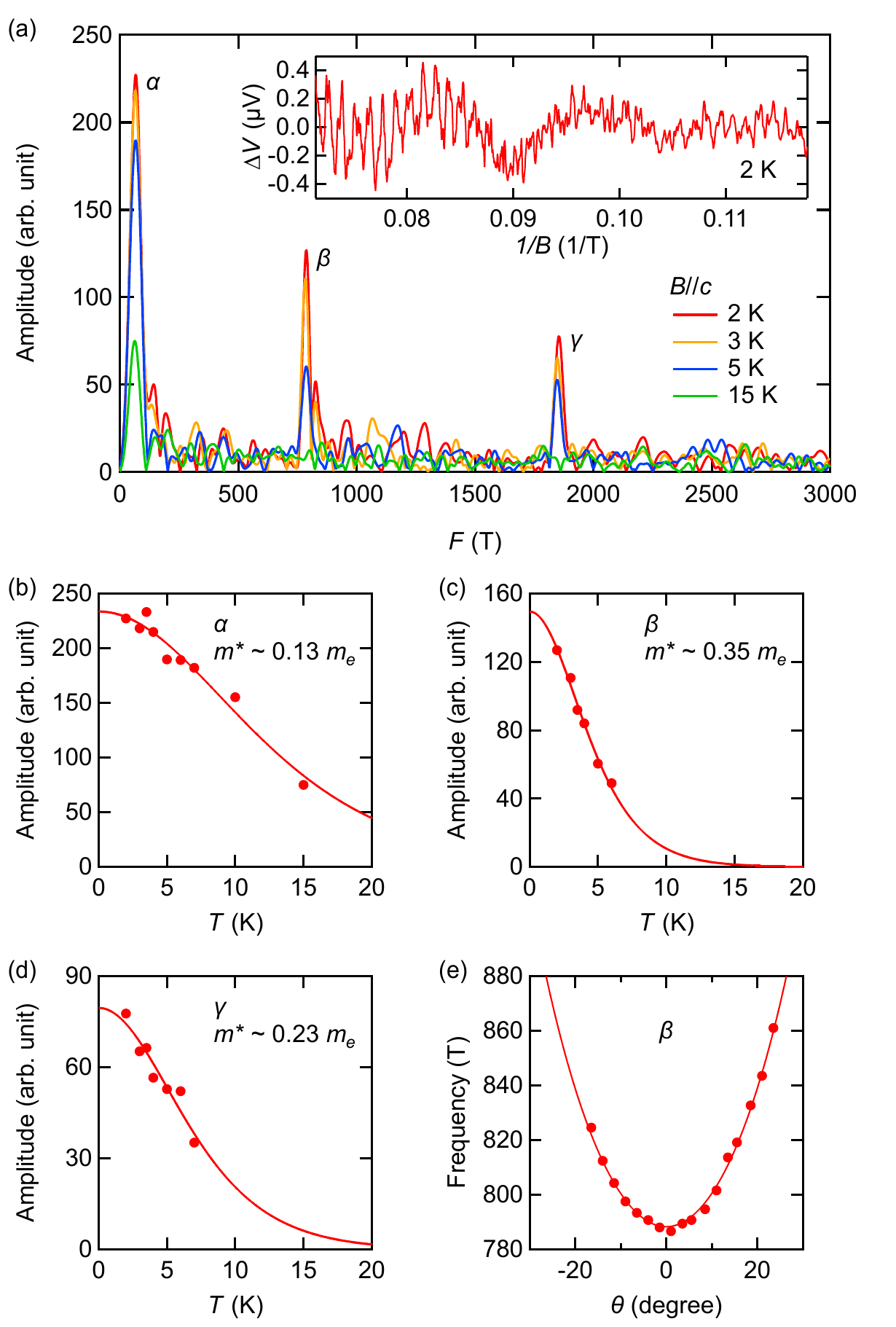}}   
              \caption{\label{fig3}
SdH quantum oscillation data. (a) FFT spectra of oscillations at different temperatures. (b)$-$(d) Temperature dependence of the oscillation amplitudes of the detected $\alpha$, $\beta$, and $\gamma$ peaks. The solid curves are the fittings using the thermal damping factor of the Lifshitz-Kosevich (LK) theory. (e) Angular dependence of the oscillation frequency of the $\beta$ peak. The data is fitted by $F(\theta)=F(0^\circ)/\cos\theta$.}
\end{figure}

The superior magnetotransport signals motivate the search of SdH quantum oscillations. To achieve this goal, we measure the MR of another sample with a slightly larger RRR (sample B1, RRR=303). The transport current is flowing along the crystalline $b$ axis in B1, and the magnetic field can be varied from $c$ axis ($\theta=0^\circ$) to $a$ axis ($\theta=90^\circ$). We also measured sample A1, in which the current direction is along the $a$ axis, as presented in the Supplemental Material \cite{SUPP}. Clear SdH oscillations can be seen when the MR background is removed via a low-order polynomial fitting, as displayed in the inset of Fig.~\ref{fig3}(a) for the dataset at 2~K. The main panel of Fig.~\ref{fig3}(a) displays the Fast Fourier Transform (FFT) spectra of the oscillations at several temperatures. Three peaks, labelled as $\alpha$, $\beta$ and $\gamma$, can be clearly identified and they correspond to frequencies 65~T, 787~T and 1852~T, respectively. Fitting the amplitudes of a peak at different temperatures by the thermal damping factor of Lifshitz-Kosevich (LK) formula \cite{Shoenberg1984Book} gives the effective mass. Figures~\ref{fig3}(b), (c) and (d) display the results of the analysis, and the resultant effective masses are 0.13~$m_e$, 0.35~$m_e$ and 0.23~$m_e$ for $\alpha$, $\beta$ and $\gamma$, respectively ($m_e$ is the rest mass of an electron). These $m^*$ values are noticeably smaller than $m_e$, indicating a weak correlation effect.  

Varying the field angle $(\theta)$ from $c$ axis while maintaining $B\,\perp \, I$, we have measured the SdH oscillations. The signal quality of $\beta$ remained robust as we increased $\theta$. From Fig.~\ref{fig3}(e), which shows the angular dependence of the SdH frequency of $\beta$, we see the experimental data (symbols) follow a $1/\cos\theta$ dependence accurately. Therefore, $\beta$ comes from a quasi-2D, cylindrical Fermi surface with the cylindrical axis parallel to the sample $c$ axis.

\begin{figure}[!t]\centering
      \resizebox{9cm}{!}{
\includegraphics{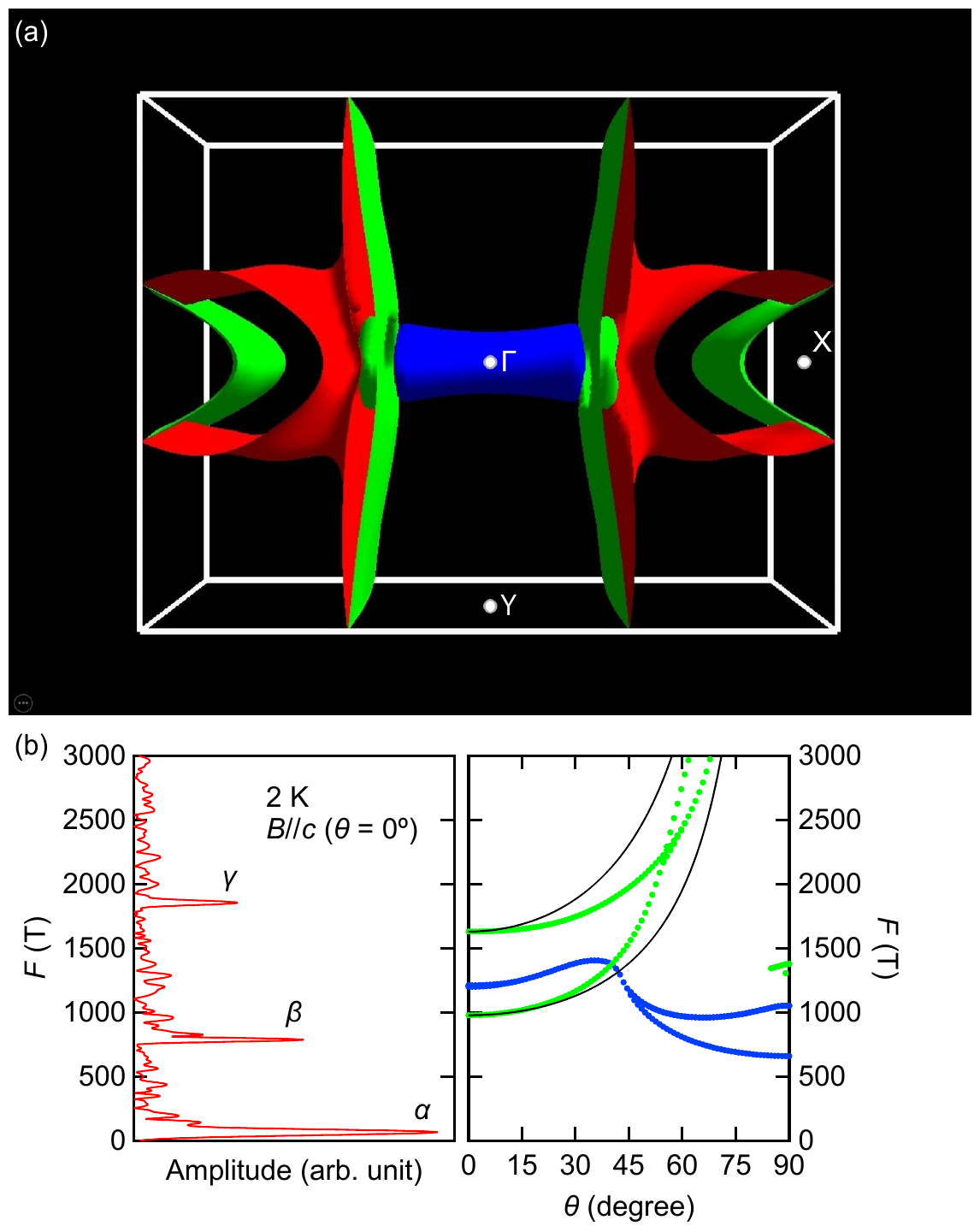}}\caption{\label{fig4} 
(a) The calculated Fermi surface of the non-CDW phase of CuTe in the first brillouin zone. (b) (right panel) The calculated angular dependence of the quantum oscillation frequencies. The colours of the curves correspond to the colours of the Fermi surface sheets shown in (a). Solid lines are the $F(\theta)=F(0^\circ)/\cos\theta$ fittings. (left panel) The FFT of the measured SdH oscillation with $B\,\parallelsum \, c$ ($\theta = 0^\circ$) at 2~K.}
\end{figure}

Figure~\ref{fig4}(a) shows the calculated Fermi surface of CuTe in the pristine phase, as viewed from the $k_z$ axis. Quantum oscillation frequencies collected with $B\,\parallelsum \, c$ are proportional to the area of extremal orbits parallel to the $k_x$-$k_y$ plane. Thus, Fig.~\ref{fig4}(a) provides a suitable visualisation for the experimental configuration with $B\,\parallelsum \, c$. At the $\Gamma$ point, there is a cylinder-like Fermi surface that extends along $k_x$. On its two sides, two sheetlike Fermi surfaces parallel to the $k_y$-$k_z$ plane cover the entire Brillouin zone. Extending from the sheets towards the faces of the Brillouin zone are two quasi-2D cylindrical Fermi surfaces centred around the X point. These features are consistent with the band structure, as well as the Fermi surface measured and calculated by Zhang \etal~\cite{Zhang2018CuTe}. They are also consistent with the calculations done by Kim \etal~\cite{Kim2019CuTe} and Cudazzo \etal~\cite{Cudazzo2021CuTe}.

In the right panel of Fig.~\ref{fig4}(b), we show the quantum oscillation frequencies as a function of $\theta$ (\ie from $B\,\parallelsum \, c$ to $B\,\parallelsum \, a$) extracted from calculation. The experimental data, identical to the 2~K spectrum displayed in Fig.~\ref{fig3}(a) are shown next to the calculated results for comparison. The calculation indicates that one band would contribute to two peaks at about 1000~T and 1600~T for $\theta = 0^\circ$ while another band would contribute to a peak at about 1200~T for $\theta = 0^\circ$. Here we only show the calculated angular dependence of the oscillation frequencies below 3000~T. The full frequency range can be found in the Supplemental Material \cite{SUPP}, where a frequency as large as 14035~T can be seen for $\theta = 0^\circ$.  We see that there are two frequencies in the calculation that shows upturn behaviours as $\theta$ increases. One of them may correspond to the measured $\beta$ peak. However, both calculated frequencies do not follow $F(\theta)=F(0^\circ)/\cos\theta$ (see solid lines in the right panel of Fig.~\ref{fig4}(b)), in contrast to the measured $\beta$ peak which follows the $1/\cos\theta$ dependence accurately. Furthermore, there is no low-frequency peak in the calculation to match the $\alpha$ peak in our measurement. Thus, the experimental data do not agree with the calculation. This is not surprising because the calculations were performed for the non-CDW phase while the measured quantum oscillations come from the CDW state. Therefore, the discrepancies support a CDW-induced Fermi surface reconstruction in CuTe. 

To take CDW into consideration for the band structure calculation, Ref.~\cite{Kim2019CuTe} argues that the Coulomb correlation must be taken into account, necessitating the usage of DFT + $U$. However, our $m^*$ values for the detected frequencies are all significantly smaller than $m_e$. Thus, a comprehensive comparison between the experimental data and calculations is likely to be a complicated issue. It is conceivable that SdH frequencies with higher effective masses have not been detected. Therefore, as an extension of the present work, quantum oscillation measurements at a lower temperature or higher magnetic field will be useful, while a series of DFT + $U$ calculations can be performed with different $U$ values. Such a dual-track approach could shed light on the role of correlation in the CDW state of CuTe, and enable a complete determination of the electronic structure of CuTe.

\section{IV.~CONCLUSION}
In conclusion, we measured the magnetoresistance of CuTe. Along the crystallographic direction $a$, the magnetoresistance tends to saturate as magnetic field increases. Along the crystallographic direction $b$, however, the magnetoresistance shows no sign of saturation. The magnetoresistance along $a$ does not obey Kohler scaling while the magnetoresistance along $b$ does. Therefore, the magnetotransport properties in CuTe is sensitive to the direction of the applied current. We also successfully detected the Shubnikov–de Haas quantum oscillations of CuTe at different temperatures and different field angles, facilitating the study of the fermiology of CuTe. Comparing the calculated band structure and our quantum oscillation data, we infer a CDW-induced Fermi surface reconstruction in CuTe.

\begin{acknowledgments}
\section{ACKNOWLEDGEMENTS}
This work was supported by Research Grants Council of Hong Kong (CUHK14300419 and A-CUHK402/19), CUHK Direct Grant (4053408, 4053461, and 4053577), Taiwan Consortium of Emergent Crystalline Materials, National Science and Technology Council of Taiwan (109-2112-M-006-013, 110-2124-M-006-006, 110-2124-M-006-010, and 110-2112-M-032-014-MY3), the National Natural Science Foundation of China (12104384 and 12174175), the Shenzhen Basic Research Fund (20220815101116001) and City University of Hong Kong (9610438, 7005610, and 9680320). We acknowledge Kai Ham Yu of The Chinese University of Hong Kong for experimental support and we are grateful to L. J. Chang of National Cheng Kung University for the help with operating the Laue diffraction spectrometer. We also thank the National Center for High-Performance Computing (NCHC) in Taiwan for providing computational and storage resources.

\end{acknowledgments}

\providecommand{\noopsort}[1]{}\providecommand{\singleletter}[1]{#1}%


\begin{thebibliography}{34}%
\makeatletter
\providecommand \@ifxundefined [1]{%
 \@ifx{#1\undefined}
}%
\providecommand \@ifnum [1]{%
 \ifnum #1\expandafter \@firstoftwo
 \else \expandafter \@secondoftwo
 \fi
}%
\providecommand \@ifx [1]{%
 \ifx #1\expandafter \@firstoftwo
 \else \expandafter \@secondoftwo
 \fi
}%
\providecommand \natexlab [1]{#1}%
\providecommand \enquote  [1]{``#1''}%
\providecommand \bibnamefont  [1]{#1}%
\providecommand \bibfnamefont [1]{#1}%
\providecommand \citenamefont [1]{#1}%
\providecommand \href@noop [0]{\@secondoftwo}%
\providecommand \href [0]{\begingroup \@sanitize@url \@href}%
\providecommand \@href[1]{\@@startlink{#1}\@@href}%
\providecommand \@@href[1]{\endgroup#1\@@endlink}%
\providecommand \@sanitize@url [0]{\catcode `\\12\catcode `\$12\catcode
  `\&12\catcode `\#12\catcode `\^12\catcode `\_12\catcode `\%12\relax}%
\providecommand \@@startlink[1]{}%
\providecommand \@@endlink[0]{}%
\providecommand \url  [0]{\begingroup\@sanitize@url \@url }%
\providecommand \@url [1]{\endgroup\@href {#1}{\urlprefix }}%
\providecommand \urlprefix  [0]{URL }%
\providecommand \Eprint [0]{\href }%
\providecommand \doibase [0]{http://dx.doi.org/}%
\providecommand \selectlanguage [0]{\@gobble}%
\providecommand \bibinfo  [0]{\@secondoftwo}%
\providecommand \bibfield  [0]{\@secondoftwo}%
\providecommand \translation [1]{[#1]}%
\providecommand \BibitemOpen [0]{}%
\providecommand \bibitemStop [0]{}%
\providecommand \bibitemNoStop [0]{.\EOS\space}%
\providecommand \EOS [0]{\spacefactor3000\relax}%
\providecommand \BibitemShut  [1]{\csname bibitem#1\endcsname}%
\let\auto@bib@innerbib\@empty
\bibitem [{\citenamefont {Gr{\"u}ner}(2018)}]{Gruner2018Book}%
  \BibitemOpen
  \bibfield  {author} {\bibinfo {author} {\bibfnamefont {G.}~\bibnamefont
  {Gr{\"u}ner}},\ }\href@noop {} {\emph {\bibinfo {title} {Density waves in
  solids}}}\ (\bibinfo  {publisher} {CRC press, Boca Raton},\ \bibinfo {year}
  {2018})\BibitemShut {NoStop}%
\bibitem [{\citenamefont {Chang}\ \emph {et~al.}(2012)\citenamefont {Chang},
  \citenamefont {Blackburn}, \citenamefont {Holmes}, \citenamefont
  {Christensen}, \citenamefont {Larsen}, \citenamefont {Mesot}, \citenamefont
  {Liang}, \citenamefont {Bonn}, \citenamefont {Hardy}, \citenamefont
  {Watenphul}, \citenamefont {Zimmermann}, \citenamefont {Forgan},\ and\
  \citenamefont {Hayden}}]{Chang2012YBCO}%
  \BibitemOpen
  \bibfield  {author} {\bibinfo {author} {\bibfnamefont {J.}~\bibnamefont
  {Chang}}, \bibinfo {author} {\bibfnamefont {E.}~\bibnamefont {Blackburn}},
  \bibinfo {author} {\bibfnamefont {A.}~\bibnamefont {Holmes}}, \bibinfo
  {author} {\bibfnamefont {N.~B.}\ \bibnamefont {Christensen}}, \bibinfo
  {author} {\bibfnamefont {J.}~\bibnamefont {Larsen}}, \bibinfo {author}
  {\bibfnamefont {J.}~\bibnamefont {Mesot}}, \bibinfo {author} {\bibfnamefont
  {R.}~\bibnamefont {Liang}}, \bibinfo {author} {\bibfnamefont
  {D.}~\bibnamefont {Bonn}}, \bibinfo {author} {\bibfnamefont {W.}~\bibnamefont
  {Hardy}}, \bibinfo {author} {\bibfnamefont {A.}~\bibnamefont {Watenphul}},
  \bibinfo {author} {\bibfnamefont {M.~v.}\ \bibnamefont {Zimmermann}},
  \bibinfo {author} {\bibfnamefont {E.~M.}\ \bibnamefont {Forgan}}, \ and\
  \bibinfo {author} {\bibfnamefont {S.~M.}\ \bibnamefont {Hayden}},\ }\bibfield
   {title} {\enquote {\bibinfo {title} {{Direct observation of competition
  between superconductivity and charge density wave order in
  {YBa$_2$Cu$_3$O$_{6.67}$}}},}\ }\href@noop {} {\bibfield  {journal} {\bibinfo
   {journal} {Nat. Phys.}\ }\textbf {\bibinfo {volume} {8}},\ \bibinfo {pages}
  {871} (\bibinfo {year} {2012})}\BibitemShut {NoStop}%
\bibitem [{\citenamefont {Dreher}\ \emph {et~al.}(2021)\citenamefont {Dreher},
  \citenamefont {Wan}, \citenamefont {Chikina}, \citenamefont {Bianchi},
  \citenamefont {Guo}, \citenamefont {Harsh}, \citenamefont
  {Ma$\tilde{n}$as-Valero}, \citenamefont {Coronado}, \citenamefont
  {Martínez-Galera}, \citenamefont {Hofmann}, \citenamefont {Miwa},\ and\
  \citenamefont {Ugeda}}]{Dreher2021NbSe2}%
  \BibitemOpen
  \bibfield  {author} {\bibinfo {author} {\bibfnamefont {P.}~\bibnamefont
  {Dreher}}, \bibinfo {author} {\bibfnamefont {W.}~\bibnamefont {Wan}},
  \bibinfo {author} {\bibfnamefont {A.}~\bibnamefont {Chikina}}, \bibinfo
  {author} {\bibfnamefont {M.}~\bibnamefont {Bianchi}}, \bibinfo {author}
  {\bibfnamefont {H.}~\bibnamefont {Guo}}, \bibinfo {author} {\bibfnamefont
  {R.}~\bibnamefont {Harsh}}, \bibinfo {author} {\bibfnamefont
  {S.}~\bibnamefont {Ma$\tilde{n}$as-Valero}}, \bibinfo {author} {\bibfnamefont
  {E.}~\bibnamefont {Coronado}}, \bibinfo {author} {\bibfnamefont {A.~J.}\
  \bibnamefont {Martínez-Galera}}, \bibinfo {author} {\bibfnamefont
  {P.}~\bibnamefont {Hofmann}}, \bibinfo {author} {\bibfnamefont {J.~A.}\
  \bibnamefont {Miwa}}, \ and\ \bibinfo {author} {\bibfnamefont {M.~M.}\
  \bibnamefont {Ugeda}},\ }\bibfield  {title} {\enquote {\bibinfo {title}
  {{Proximity effects on the charge density wave order and superconductivity in
  single-layer NbSe$_2$}},}\ }\href@noop {} {\bibfield  {journal} {\bibinfo
  {journal} {ACS Nano}\ }\textbf {\bibinfo {volume} {15}},\ \bibinfo {pages}
  {19430} (\bibinfo {year} {2021})}\BibitemShut {NoStop}%
\bibitem [{\citenamefont {Kusmartseva}\ \emph {et~al.}(2009)\citenamefont
  {Kusmartseva}, \citenamefont {Sipos}, \citenamefont {Berger}, \citenamefont
  {Forro},\ and\ \citenamefont {Tuti{\v{s}}}}]{Kusmartseva2009TiSe2}%
  \BibitemOpen
  \bibfield  {author} {\bibinfo {author} {\bibfnamefont {A.~F.}\ \bibnamefont
  {Kusmartseva}}, \bibinfo {author} {\bibfnamefont {B.}~\bibnamefont {Sipos}},
  \bibinfo {author} {\bibfnamefont {H.}~\bibnamefont {Berger}}, \bibinfo
  {author} {\bibfnamefont {L.}~\bibnamefont {Forro}}, \ and\ \bibinfo {author}
  {\bibfnamefont {E.}~\bibnamefont {Tuti{\v{s}}}},\ }\bibfield  {title}
  {\enquote {\bibinfo {title} {{Pressure Induced Superconductivity in Pristine
  1T-TiSe$_2$}},}\ }\href@noop {} {\bibfield  {journal} {\bibinfo  {journal}
  {Phys. Rev. Lett.}\ }\textbf {\bibinfo {volume} {103}},\ \bibinfo {pages}
  {236401} (\bibinfo {year} {2009})}\BibitemShut {NoStop}%
\bibitem [{\citenamefont {Kudo}\ \emph {et~al.}(2010)\citenamefont {Kudo},
  \citenamefont {Nishikubo},\ and\ \citenamefont {Nohara}}]{Kudo2010SrPt2As2}%
  \BibitemOpen
  \bibfield  {author} {\bibinfo {author} {\bibfnamefont {K.}~\bibnamefont
  {Kudo}}, \bibinfo {author} {\bibfnamefont {Y.}~\bibnamefont {Nishikubo}}, \
  and\ \bibinfo {author} {\bibfnamefont {M.}~\bibnamefont {Nohara}},\
  }\bibfield  {title} {\enquote {\bibinfo {title} {{Coexistence of
  superconductivity and charge density wave in SrPt$_2$As$_2$}},}\ }\href@noop
  {} {\bibfield  {journal} {\bibinfo  {journal} {J. Phys. Soc. Jpn.}\ }\textbf
  {\bibinfo {volume} {79}},\ \bibinfo {pages} {123710} (\bibinfo {year}
  {2010})}\BibitemShut {NoStop}%
\bibitem [{\citenamefont {Li}\ \emph {et~al.}(2017)\citenamefont {Li},
  \citenamefont {Zhu}, \citenamefont {Stewart},\ and\ \citenamefont
  {Hebard}}]{Li2017TaS2}%
  \BibitemOpen
  \bibfield  {author} {\bibinfo {author} {\bibfnamefont {A.~J.}\ \bibnamefont
  {Li}}, \bibinfo {author} {\bibfnamefont {X.}~\bibnamefont {Zhu}}, \bibinfo
  {author} {\bibfnamefont {G.}~\bibnamefont {Stewart}}, \ and\ \bibinfo
  {author} {\bibfnamefont {A.~F.}\ \bibnamefont {Hebard}},\ }\bibfield  {title}
  {\enquote {\bibinfo {title} {{Bi-2212/1T-TaS$_2$ Van der Waals junctions:
  Interplay of proximity induced high-T$_c$ superconductivity and CDW
  order}},}\ }\href@noop {} {\bibfield  {journal} {\bibinfo  {journal} {Sci.
  Rep.}\ }\textbf {\bibinfo {volume} {7}},\ \bibinfo {pages} {4639} (\bibinfo
  {year} {2017})}\BibitemShut {NoStop}%
\bibitem [{\citenamefont {Chen}\ \emph {et~al.}(2021)\citenamefont {Chen},
  \citenamefont {Wang}, \citenamefont {Yin}, \citenamefont {Gu}, \citenamefont
  {Jiang}, \citenamefont {Tu}, \citenamefont {Gong}, \citenamefont {Uwatoko},
  \citenamefont {Sun}, \citenamefont {Lei}, \citenamefont {Hu},\ and\
  \citenamefont {Cheng}}]{Chen2021CsV3Sb5}%
  \BibitemOpen
  \bibfield  {author} {\bibinfo {author} {\bibfnamefont {K.~Y.}\ \bibnamefont
  {Chen}}, \bibinfo {author} {\bibfnamefont {N.~N.}\ \bibnamefont {Wang}},
  \bibinfo {author} {\bibfnamefont {Q.~W.}\ \bibnamefont {Yin}}, \bibinfo
  {author} {\bibfnamefont {Y.~H.}\ \bibnamefont {Gu}}, \bibinfo {author}
  {\bibfnamefont {K.}~\bibnamefont {Jiang}}, \bibinfo {author} {\bibfnamefont
  {Z.~J.}\ \bibnamefont {Tu}}, \bibinfo {author} {\bibfnamefont {C.~S.}\
  \bibnamefont {Gong}}, \bibinfo {author} {\bibfnamefont {Y.}~\bibnamefont
  {Uwatoko}}, \bibinfo {author} {\bibfnamefont {J.~P.}\ \bibnamefont {Sun}},
  \bibinfo {author} {\bibfnamefont {H.~C.}\ \bibnamefont {Lei}}, \bibinfo
  {author} {\bibfnamefont {J.~P.}\ \bibnamefont {Hu}}, \ and\ \bibinfo {author}
  {\bibfnamefont {J.~G.}\ \bibnamefont {Cheng}},\ }\bibfield  {title} {\enquote
  {\bibinfo {title} {{Double superconducting dome and triple enhancement of
  T$_c$ in the kagome superconductor CsV$_3$Sb$_5$ under high pressure}},}\
  }\href@noop {} {\bibfield  {journal} {\bibinfo  {journal} {Phys. Rev. Lett.}\
  }\textbf {\bibinfo {volume} {126}},\ \bibinfo {pages} {247001} (\bibinfo
  {year} {2021})}\BibitemShut {NoStop}%
\bibitem [{\citenamefont {Neupert}\ \emph {et~al.}(2022)\citenamefont
  {Neupert}, \citenamefont {Denner}, \citenamefont {Yin}, \citenamefont
  {Thomale},\ and\ \citenamefont {Hasan}}]{Neupert2022AV3Sb5}%
  \BibitemOpen
  \bibfield  {author} {\bibinfo {author} {\bibfnamefont {T.}~\bibnamefont
  {Neupert}}, \bibinfo {author} {\bibfnamefont {M.~M.}\ \bibnamefont {Denner}},
  \bibinfo {author} {\bibfnamefont {J.-X.}\ \bibnamefont {Yin}}, \bibinfo
  {author} {\bibfnamefont {R.}~\bibnamefont {Thomale}}, \ and\ \bibinfo
  {author} {\bibfnamefont {M.~Z.}\ \bibnamefont {Hasan}},\ }\bibfield  {title}
  {\enquote {\bibinfo {title} {{Charge order and superconductivity in kagome
  materials}},}\ }\href@noop {} {\bibfield  {journal} {\bibinfo  {journal}
  {Nat. Phys.}\ }\textbf {\bibinfo {volume} {18}},\ \bibinfo {pages} {137}
  (\bibinfo {year} {2022})}\BibitemShut {NoStop}%
\bibitem [{\citenamefont {Wang}\ \emph
  {et~al.}(2021{\natexlab{a}})\citenamefont {Wang}, \citenamefont {Kong},
  \citenamefont {Shi}, \citenamefont {Pei}, \citenamefont {Wen}, \citenamefont
  {Gao}, \citenamefont {Zhao}, \citenamefont {Yin}, \citenamefont {Wu},
  \citenamefont {Li}, \citenamefont {Lei}, \citenamefont {Li}, \citenamefont
  {Chen}, \citenamefont {Yan},\ and\ \citenamefont {Qi}}]{Wang2021CsV3Sb5}%
  \BibitemOpen
  \bibfield  {author} {\bibinfo {author} {\bibfnamefont {Q.}~\bibnamefont
  {Wang}}, \bibinfo {author} {\bibfnamefont {P.}~\bibnamefont {Kong}}, \bibinfo
  {author} {\bibfnamefont {W.}~\bibnamefont {Shi}}, \bibinfo {author}
  {\bibfnamefont {C.}~\bibnamefont {Pei}}, \bibinfo {author} {\bibfnamefont
  {C.}~\bibnamefont {Wen}}, \bibinfo {author} {\bibfnamefont {L.}~\bibnamefont
  {Gao}}, \bibinfo {author} {\bibfnamefont {Y.}~\bibnamefont {Zhao}}, \bibinfo
  {author} {\bibfnamefont {Q.}~\bibnamefont {Yin}}, \bibinfo {author}
  {\bibfnamefont {Y.}~\bibnamefont {Wu}}, \bibinfo {author} {\bibfnamefont
  {G.}~\bibnamefont {Li}}, \bibinfo {author} {\bibfnamefont {H.}~\bibnamefont
  {Lei}}, \bibinfo {author} {\bibfnamefont {J.}~\bibnamefont {Li}}, \bibinfo
  {author} {\bibfnamefont {Y.}~\bibnamefont {Chen}}, \bibinfo {author}
  {\bibfnamefont {S.}~\bibnamefont {Yan}}, \ and\ \bibinfo {author}
  {\bibfnamefont {Y.}~\bibnamefont {Qi}},\ }\bibfield  {title} {\enquote
  {\bibinfo {title} {{Charge density wave orders and enhanced superconductivity
  under pressure in the kagome metal CsV$_3$Sb$_5$}},}\ }\href@noop {}
  {\bibfield  {journal} {\bibinfo  {journal} {Adv. Mater.}\ }\textbf {\bibinfo
  {volume} {33}},\ \bibinfo {pages} {2102813} (\bibinfo {year}
  {2021}{\natexlab{a}})}\BibitemShut {NoStop}%
\bibitem [{\citenamefont {Sipos}\ \emph {et~al.}(2008)\citenamefont {Sipos},
  \citenamefont {Kusmartseva}, \citenamefont {Akrap}, \citenamefont {Berger},
  \citenamefont {Forr{\'o}},\ and\ \citenamefont
  {Tuti{\v{s}}}}]{Sipos2008TaSe2}%
  \BibitemOpen
  \bibfield  {author} {\bibinfo {author} {\bibfnamefont {B.}~\bibnamefont
  {Sipos}}, \bibinfo {author} {\bibfnamefont {A.~F.}\ \bibnamefont
  {Kusmartseva}}, \bibinfo {author} {\bibfnamefont {A.}~\bibnamefont {Akrap}},
  \bibinfo {author} {\bibfnamefont {H.}~\bibnamefont {Berger}}, \bibinfo
  {author} {\bibfnamefont {L.}~\bibnamefont {Forr{\'o}}}, \ and\ \bibinfo
  {author} {\bibfnamefont {E.}~\bibnamefont {Tuti{\v{s}}}},\ }\bibfield
  {title} {\enquote {\bibinfo {title} {{From Mott state to superconductivity in
  1T-TaS$_2$}},}\ }\href@noop {} {\bibfield  {journal} {\bibinfo  {journal}
  {Nat. Mater.}\ }\textbf {\bibinfo {volume} {7}},\ \bibinfo {pages} {960}
  (\bibinfo {year} {2008})}\BibitemShut {NoStop}%
\bibitem [{\citenamefont {Liu}\ \emph {et~al.}(2016)\citenamefont {Liu},
  \citenamefont {Shao}, \citenamefont {Li}, \citenamefont {Lu}, \citenamefont
  {Zhu}, \citenamefont {Tong}, \citenamefont {Xiao}, \citenamefont {Ling},
  \citenamefont {Xi}, \citenamefont {Pi}, \citenamefont {Tian}, \citenamefont
  {Yang}, \citenamefont {Li}, \citenamefont {Song}, \citenamefont {Zhu},\ and\
  \citenamefont {Sun}}]{Liu2016TaSe2}%
  \BibitemOpen
  \bibfield  {author} {\bibinfo {author} {\bibfnamefont {Y.}~\bibnamefont
  {Liu}}, \bibinfo {author} {\bibfnamefont {D.~F.}\ \bibnamefont {Shao}},
  \bibinfo {author} {\bibfnamefont {L.~J.}\ \bibnamefont {Li}}, \bibinfo
  {author} {\bibfnamefont {W.~J.}\ \bibnamefont {Lu}}, \bibinfo {author}
  {\bibfnamefont {X.~D.}\ \bibnamefont {Zhu}}, \bibinfo {author} {\bibfnamefont
  {P.}~\bibnamefont {Tong}}, \bibinfo {author} {\bibfnamefont {R.~C.}\
  \bibnamefont {Xiao}}, \bibinfo {author} {\bibfnamefont {L.~S.}\ \bibnamefont
  {Ling}}, \bibinfo {author} {\bibfnamefont {C.~Y.}\ \bibnamefont {Xi}},
  \bibinfo {author} {\bibfnamefont {L.}~\bibnamefont {Pi}}, \bibinfo {author}
  {\bibfnamefont {H.~F.}\ \bibnamefont {Tian}}, \bibinfo {author}
  {\bibfnamefont {H.~X.}\ \bibnamefont {Yang}}, \bibinfo {author}
  {\bibfnamefont {J.~Q.}\ \bibnamefont {Li}}, \bibinfo {author} {\bibfnamefont
  {W.~H.}\ \bibnamefont {Song}}, \bibinfo {author} {\bibfnamefont {X.~B.}\
  \bibnamefont {Zhu}}, \ and\ \bibinfo {author} {\bibfnamefont {Y.~P.}\
  \bibnamefont {Sun}},\ }\bibfield  {title} {\enquote {\bibinfo {title}
  {{Nature of charge density waves and superconductivity in
  1T-TaSe$_{2-x}$Te$_x$}},}\ }\href@noop {} {\bibfield  {journal} {\bibinfo
  {journal} {Phys. Rev. B}\ }\textbf {\bibinfo {volume} {94}},\ \bibinfo
  {pages} {045131} (\bibinfo {year} {2016})}\BibitemShut {NoStop}%
\bibitem [{\citenamefont {Stolze}\ \emph {et~al.}(2013)\citenamefont {Stolze},
  \citenamefont {Isaeva}, \citenamefont {Nitsche}, \citenamefont {Burkhardt},
  \citenamefont {Lichte}, \citenamefont {Wolf},\ and\ \citenamefont
  {Doert}}]{Stolze2013CuTe}%
  \BibitemOpen
  \bibfield  {author} {\bibinfo {author} {\bibfnamefont {K.}~\bibnamefont
  {Stolze}}, \bibinfo {author} {\bibfnamefont {A.}~\bibnamefont {Isaeva}},
  \bibinfo {author} {\bibfnamefont {F.}~\bibnamefont {Nitsche}}, \bibinfo
  {author} {\bibfnamefont {U.}~\bibnamefont {Burkhardt}}, \bibinfo {author}
  {\bibfnamefont {H.}~\bibnamefont {Lichte}}, \bibinfo {author} {\bibfnamefont
  {D.}~\bibnamefont {Wolf}}, \ and\ \bibinfo {author} {\bibfnamefont
  {T.}~\bibnamefont {Doert}},\ }\bibfield  {title} {\enquote {\bibinfo {title}
  {{CuTe: remarkable bonding features as a consequence of a charge density
  wave}},}\ }\href@noop {} {\bibfield  {journal} {\bibinfo  {journal} {Angew.
  Chem. Int. Ed.}\ }\textbf {\bibinfo {volume} {52}},\ \bibinfo {pages} {862}
  (\bibinfo {year} {2013})}\BibitemShut {NoStop}%
\bibitem [{\citenamefont {Zhang}\ \emph {et~al.}(2018)\citenamefont {Zhang},
  \citenamefont {Liu}, \citenamefont {Zhang}, \citenamefont {Deng},
  \citenamefont {Yan}, \citenamefont {Yao}, \citenamefont {Zheng},
  \citenamefont {Schwier}, \citenamefont {Shimada}, \citenamefont {Denlinger},
  \citenamefont {Wu}, \citenamefont {Duan},\ and\ \citenamefont
  {Zhou}}]{Zhang2018CuTe}%
  \BibitemOpen
  \bibfield  {author} {\bibinfo {author} {\bibfnamefont {K.}~\bibnamefont
  {Zhang}}, \bibinfo {author} {\bibfnamefont {X.}~\bibnamefont {Liu}}, \bibinfo
  {author} {\bibfnamefont {H.}~\bibnamefont {Zhang}}, \bibinfo {author}
  {\bibfnamefont {K.}~\bibnamefont {Deng}}, \bibinfo {author} {\bibfnamefont
  {M.}~\bibnamefont {Yan}}, \bibinfo {author} {\bibfnamefont {W.}~\bibnamefont
  {Yao}}, \bibinfo {author} {\bibfnamefont {M.}~\bibnamefont {Zheng}}, \bibinfo
  {author} {\bibfnamefont {E.~F.}\ \bibnamefont {Schwier}}, \bibinfo {author}
  {\bibfnamefont {K.}~\bibnamefont {Shimada}}, \bibinfo {author} {\bibfnamefont
  {J.~D.}\ \bibnamefont {Denlinger}}, \bibinfo {author} {\bibfnamefont
  {Y.}~\bibnamefont {Wu}}, \bibinfo {author} {\bibfnamefont {W.}~\bibnamefont
  {Duan}}, \ and\ \bibinfo {author} {\bibfnamefont {S.}~\bibnamefont {Zhou}},\
  }\bibfield  {title} {\enquote {\bibinfo {title} {{Evidence for a
  quasi-one-dimensional charge density wave in CuTe by angle-resolved
  photoemission spectroscopy}},}\ }\href@noop {} {\bibfield  {journal}
  {\bibinfo  {journal} {Phys. Rev. Lett.}\ }\textbf {\bibinfo {volume} {121}},\
  \bibinfo {pages} {206402} (\bibinfo {year} {2018})}\BibitemShut {NoStop}%
\bibitem [{\citenamefont {Kuo}\ \emph {et~al.}(2020)\citenamefont {Kuo},
  \citenamefont {Huang}, \citenamefont {Kuo},\ and\ \citenamefont
  {Lue}}]{Kuo2020CuTe}%
  \BibitemOpen
  \bibfield  {author} {\bibinfo {author} {\bibfnamefont {C.}~\bibnamefont
  {Kuo}}, \bibinfo {author} {\bibfnamefont {R.}~\bibnamefont {Huang}}, \bibinfo
  {author} {\bibfnamefont {Y.}~\bibnamefont {Kuo}}, \ and\ \bibinfo {author}
  {\bibfnamefont {C.}~\bibnamefont {Lue}},\ }\bibfield  {title} {\enquote
  {\bibinfo {title} {{Transport and thermal behavior of the charge density wave
  phase transition in CuTe}},}\ }\href@noop {} {\bibfield  {journal} {\bibinfo
  {journal} {Phys. Rev. B}\ }\textbf {\bibinfo {volume} {102}},\ \bibinfo
  {pages} {155137} (\bibinfo {year} {2020})}\BibitemShut {NoStop}%
\bibitem [{\citenamefont {Wang}\ \emph
  {et~al.}(2021{\natexlab{b}})\citenamefont {Wang}, \citenamefont {Chen},
  \citenamefont {An}, \citenamefont {Zhou}, \citenamefont {Zhou}, \citenamefont
  {Gu}, \citenamefont {Zhang}, \citenamefont {Yang},\ and\ \citenamefont
  {Yang}}]{Wang2021CuTe}%
  \BibitemOpen
  \bibfield  {author} {\bibinfo {author} {\bibfnamefont {S.}~\bibnamefont
  {Wang}}, \bibinfo {author} {\bibfnamefont {X.}~\bibnamefont {Chen}}, \bibinfo
  {author} {\bibfnamefont {C.}~\bibnamefont {An}}, \bibinfo {author}
  {\bibfnamefont {Y.}~\bibnamefont {Zhou}}, \bibinfo {author} {\bibfnamefont
  {Y.}~\bibnamefont {Zhou}}, \bibinfo {author} {\bibfnamefont {C.}~\bibnamefont
  {Gu}}, \bibinfo {author} {\bibfnamefont {L.}~\bibnamefont {Zhang}}, \bibinfo
  {author} {\bibfnamefont {X.}~\bibnamefont {Yang}}, \ and\ \bibinfo {author}
  {\bibfnamefont {Z.}~\bibnamefont {Yang}},\ }\bibfield  {title} {\enquote
  {\bibinfo {title} {{Pressure-induced superconductivity in the
  quasi-one-dimensional charge density wave material CuTe}},}\ }\href@noop {}
  {\bibfield  {journal} {\bibinfo  {journal} {Phys. Rev. B}\ }\textbf {\bibinfo
  {volume} {103}},\ \bibinfo {pages} {134518} (\bibinfo {year}
  {2021}{\natexlab{b}})}\BibitemShut {NoStop}%
\bibitem [{\citenamefont {Wang}\ \emph {et~al.}(2022)\citenamefont {Wang},
  \citenamefont {Chen}, \citenamefont {An}, \citenamefont {Zhou}, \citenamefont
  {Zhang}, \citenamefont {Zhou}, \citenamefont {Han},\ and\ \citenamefont
  {Yang}}]{Wang2022CuTe}%
  \BibitemOpen
  \bibfield  {author} {\bibinfo {author} {\bibfnamefont {S.}~\bibnamefont
  {Wang}}, \bibinfo {author} {\bibfnamefont {X.}~\bibnamefont {Chen}}, \bibinfo
  {author} {\bibfnamefont {C.}~\bibnamefont {An}}, \bibinfo {author}
  {\bibfnamefont {Y.}~\bibnamefont {Zhou}}, \bibinfo {author} {\bibfnamefont
  {M.}~\bibnamefont {Zhang}}, \bibinfo {author} {\bibfnamefont
  {Y.}~\bibnamefont {Zhou}}, \bibinfo {author} {\bibfnamefont {Y.}~\bibnamefont
  {Han}}, \ and\ \bibinfo {author} {\bibfnamefont {Z.}~\bibnamefont {Yang}},\
  }\bibfield  {title} {\enquote {\bibinfo {title} {{Observation of
  room-temperature amplitude mode in quasi-one-dimensional charge-density-wave
  material CuTe}},}\ }\href@noop {} {\bibfield  {journal} {\bibinfo  {journal}
  {Appl. Phys. Lett.}\ }\textbf {\bibinfo {volume} {120}},\ \bibinfo {pages}
  {151902} (\bibinfo {year} {2022})}\BibitemShut {NoStop}%
\bibitem [{\citenamefont {Li}\ \emph {et~al.}(2022)\citenamefont {Li},
  \citenamefont {Yue}, \citenamefont {Wu}, \citenamefont {Xu}, \citenamefont
  {Liu}, \citenamefont {Wang}, \citenamefont {Hu}, \citenamefont {Zhou},
  \citenamefont {Shi}, \citenamefont {Zhang}, \citenamefont {Wu}, \citenamefont
  {Dong},\ and\ \citenamefont {Wang}}]{Li2022CuTe}%
  \BibitemOpen
  \bibfield  {author} {\bibinfo {author} {\bibfnamefont {R.~S.}\ \bibnamefont
  {Li}}, \bibinfo {author} {\bibfnamefont {L.}~\bibnamefont {Yue}}, \bibinfo
  {author} {\bibfnamefont {Q.}~\bibnamefont {Wu}}, \bibinfo {author}
  {\bibfnamefont {S.~X.}\ \bibnamefont {Xu}}, \bibinfo {author} {\bibfnamefont
  {Q.~M.}\ \bibnamefont {Liu}}, \bibinfo {author} {\bibfnamefont {Z.~X.}\
  \bibnamefont {Wang}}, \bibinfo {author} {\bibfnamefont {T.~C.}\ \bibnamefont
  {Hu}}, \bibinfo {author} {\bibfnamefont {X.~Y.}\ \bibnamefont {Zhou}},
  \bibinfo {author} {\bibfnamefont {L.~Y.}\ \bibnamefont {Shi}}, \bibinfo
  {author} {\bibfnamefont {S.~J.}\ \bibnamefont {Zhang}}, \bibinfo {author}
  {\bibfnamefont {D.}~\bibnamefont {Wu}}, \bibinfo {author} {\bibfnamefont
  {T.}~\bibnamefont {Dong}}, \ and\ \bibinfo {author} {\bibfnamefont {N.~L.}\
  \bibnamefont {Wang}},\ }\bibfield  {title} {\enquote {\bibinfo {title}
  {{Optical spectroscopy and ultrafast pump-probe study of a
  quasi-one-dimensional charge density wave in CuTe}},}\ }\href@noop {}
  {\bibfield  {journal} {\bibinfo  {journal} {Phys. Rev. B}\ }\textbf {\bibinfo
  {volume} {105}},\ \bibinfo {pages} {115102} (\bibinfo {year}
  {2022})}\BibitemShut {NoStop}%
\bibitem [{\citenamefont {Kim}\ \emph {et~al.}(2019)\citenamefont {Kim},
  \citenamefont {Kim},\ and\ \citenamefont {Kim}}]{Kim2019CuTe}%
  \BibitemOpen
  \bibfield  {author} {\bibinfo {author} {\bibfnamefont {S.}~\bibnamefont
  {Kim}}, \bibinfo {author} {\bibfnamefont {B.}~\bibnamefont {Kim}}, \ and\
  \bibinfo {author} {\bibfnamefont {K.}~\bibnamefont {Kim}},\ }\bibfield
  {title} {\enquote {\bibinfo {title} {{Role of Coulomb correlations in the
  charge density wave of CuTe}},}\ }\href@noop {} {\bibfield  {journal}
  {\bibinfo  {journal} {Phys. Rev. B}\ }\textbf {\bibinfo {volume} {100}},\
  \bibinfo {pages} {054112} (\bibinfo {year} {2019})}\BibitemShut {NoStop}%
\bibitem [{\citenamefont {Cudazzo}\ and\ \citenamefont
  {Wirtz}(2021)}]{Cudazzo2021CuTe}%
  \BibitemOpen
  \bibfield  {author} {\bibinfo {author} {\bibfnamefont {P.}~\bibnamefont
  {Cudazzo}}\ and\ \bibinfo {author} {\bibfnamefont {L.}~\bibnamefont
  {Wirtz}},\ }\bibfield  {title} {\enquote {\bibinfo {title} {{Collective
  electronic excitations in charge density wave systems: The case of CuTe}},}\
  }\href@noop {} {\bibfield  {journal} {\bibinfo  {journal} {Phys. Rev. B}\
  }\textbf {\bibinfo {volume} {104}},\ \bibinfo {pages} {125101} (\bibinfo
  {year} {2021})}\BibitemShut {NoStop}%
\bibitem [{\citenamefont {Pertlik}(2001)}]{Pertlik2001CuTe}%
  \BibitemOpen
  \bibfield  {author} {\bibinfo {author} {\bibfnamefont {F.}~\bibnamefont
  {Pertlik}},\ }\bibfield  {title} {\enquote {\bibinfo {title} {Vulcanite,
  {CuTe}: hydrothermal synthesis and crystal structure refinement},}\
  }\href@noop {} {\bibfield  {journal} {\bibinfo  {journal} {Mineral Petrol}\
  }\textbf {\bibinfo {volume} {71}},\ \bibinfo {pages} {149} (\bibinfo {year}
  {2001})}\BibitemShut {NoStop}%
\bibitem [{\citenamefont {Seong}\ \emph {et~al.}(1994)\citenamefont {Seong},
  \citenamefont {Albright}, \citenamefont {Zhang},\ and\ \citenamefont
  {Kanatzidis}}]{Seong1994CuTe}%
  \BibitemOpen
  \bibfield  {author} {\bibinfo {author} {\bibfnamefont {S.}~\bibnamefont
  {Seong}}, \bibinfo {author} {\bibfnamefont {T.~A.}\ \bibnamefont {Albright}},
  \bibinfo {author} {\bibfnamefont {X.}~\bibnamefont {Zhang}}, \ and\ \bibinfo
  {author} {\bibfnamefont {M.}~\bibnamefont {Kanatzidis}},\ }\bibfield  {title}
  {\enquote {\bibinfo {title} {{{Te}-{Te} bonding in copper tellurides}},}\
  }\href@noop {} {\bibfield  {journal} {\bibinfo  {journal} {J. Am. Chem.
  Soc.}\ }\textbf {\bibinfo {volume} {116}},\ \bibinfo {pages} {7287} (\bibinfo
  {year} {1994})}\BibitemShut {NoStop}%
\bibitem [{\citenamefont {Wang}\ \emph {et~al.}(2023)\citenamefont {Wang},
  \citenamefont {Wang}, \citenamefont {An}, \citenamefont {Zhou}, \citenamefont
  {Zhou}, \citenamefont {Chen}, \citenamefont {Hao},\ and\ \citenamefont
  {Yang}}]{Wang2022CuTeb}%
  \BibitemOpen
  \bibfield  {author} {\bibinfo {author} {\bibfnamefont {S.}~\bibnamefont
  {Wang}}, \bibinfo {author} {\bibfnamefont {Q.}~\bibnamefont {Wang}}, \bibinfo
  {author} {\bibfnamefont {C.}~\bibnamefont {An}}, \bibinfo {author}
  {\bibfnamefont {Y.}~\bibnamefont {Zhou}}, \bibinfo {author} {\bibfnamefont
  {Y.}~\bibnamefont {Zhou}}, \bibinfo {author} {\bibfnamefont {X.}~\bibnamefont
  {Chen}}, \bibinfo {author} {\bibfnamefont {N.}~\bibnamefont {Hao}}, \ and\
  \bibinfo {author} {\bibfnamefont {Z.}~\bibnamefont {Yang}},\ }\bibfield
  {title} {\enquote {\bibinfo {title} {{Two distinct charge density wave orders
  and emergent superconductivity in pressurized CuTe}},}\ }\href {\doibase
  https://doi.org/10.1016/j.matt.2023.07.018} {\bibfield  {journal} {\bibinfo
  {journal} {Matter}\ } (\bibinfo {year} {2023}),\
  https://doi.org/10.1016/j.matt.2023.07.018}\BibitemShut {NoStop}%
\bibitem [{Foo()}]{Footnote1}%
  \BibitemOpen
  \href@noop {} {}\bibinfo {note} {A synchrotron X-ray powder diffraction
  (SXRD) experiment was conducted at room temperature for phase identification
  and structure analysis. The sample was packed in a 0.1-mm-diameter
  borosilicate capillary, which was kept spinning during data collection. The
  SXRD pattern was collected with the MYTHEN detector with 15 keV beam at
  Beamline 19A, Taiwan Photon Source, National Synchrotron Radiation Research
  Center (NSRRC) in Taiwan.}\BibitemShut {Stop}%
\bibitem [{\citenamefont {Giannozzi}\ \emph {et~al.}(2020)\citenamefont
  {Giannozzi}, \citenamefont {Baseggio}, \citenamefont {Bonf{\`a}},
  \citenamefont {Brunato}, \citenamefont {Car}, \citenamefont {Carnimeo},
  \citenamefont {Cavazzoni}, \citenamefont {De~Gironcoli}, \citenamefont
  {Delugas}, \citenamefont {Ferrari~Ruffino}, \citenamefont {Ferretti},
  \citenamefont {Marzari}, \citenamefont {Timrov}, \citenamefont {Urru},\ and\
  \citenamefont {Baroni}}]{Giannozzi2020QuantumExpresso}%
  \BibitemOpen
  \bibfield  {author} {\bibinfo {author} {\bibfnamefont {P.}~\bibnamefont
  {Giannozzi}}, \bibinfo {author} {\bibfnamefont {O.}~\bibnamefont {Baseggio}},
  \bibinfo {author} {\bibfnamefont {P.}~\bibnamefont {Bonf{\`a}}}, \bibinfo
  {author} {\bibfnamefont {D.}~\bibnamefont {Brunato}}, \bibinfo {author}
  {\bibfnamefont {R.}~\bibnamefont {Car}}, \bibinfo {author} {\bibfnamefont
  {I.}~\bibnamefont {Carnimeo}}, \bibinfo {author} {\bibfnamefont
  {C.}~\bibnamefont {Cavazzoni}}, \bibinfo {author} {\bibfnamefont
  {S.}~\bibnamefont {De~Gironcoli}}, \bibinfo {author} {\bibfnamefont
  {P.}~\bibnamefont {Delugas}}, \bibinfo {author} {\bibfnamefont
  {F.}~\bibnamefont {Ferrari~Ruffino}}, \bibinfo {author} {\bibfnamefont
  {A.}~\bibnamefont {Ferretti}}, \bibinfo {author} {\bibfnamefont
  {N.}~\bibnamefont {Marzari}}, \bibinfo {author} {\bibfnamefont
  {I.}~\bibnamefont {Timrov}}, \bibinfo {author} {\bibfnamefont
  {A.}~\bibnamefont {Urru}}, \ and\ \bibinfo {author} {\bibfnamefont
  {S.}~\bibnamefont {Baroni}},\ }\bibfield  {title} {\enquote {\bibinfo {title}
  {{Quantum ESPRESSO toward the exascale}},}\ }\href@noop {} {\bibfield
  {journal} {\bibinfo  {journal} {J. Chem. Phys.}\ }\textbf {\bibinfo {volume}
  {152}},\ \bibinfo {pages} {154105} (\bibinfo {year} {2020})}\BibitemShut
  {NoStop}%
\bibitem [{\citenamefont {Kawamura}(2019)}]{Kawamura2019FermiSurfer}%
  \BibitemOpen
  \bibfield  {author} {\bibinfo {author} {\bibfnamefont {M.}~\bibnamefont
  {Kawamura}},\ }\bibfield  {title} {\enquote {\bibinfo {title} {{FermiSurfer:
  Fermi-surface viewer providing multiple representation schemes}},}\
  }\href@noop {} {\bibfield  {journal} {\bibinfo  {journal} {Comput. Phys.
  Commun.}\ }\textbf {\bibinfo {volume} {239}},\ \bibinfo {pages} {197}
  (\bibinfo {year} {2019})}\BibitemShut {NoStop}%
\bibitem [{\citenamefont {Rourke}\ and\ \citenamefont
  {Julian}(2012)}]{Rourke2012SKEAF}%
  \BibitemOpen
  \bibfield  {author} {\bibinfo {author} {\bibfnamefont {P.}~\bibnamefont
  {Rourke}}\ and\ \bibinfo {author} {\bibfnamefont {S.}~\bibnamefont
  {Julian}},\ }\bibfield  {title} {\enquote {\bibinfo {title} {{Numerical
  extraction of de Haas--van Alphen frequencies from calculated band
  energies}},}\ }\href@noop {} {\bibfield  {journal} {\bibinfo  {journal}
  {Comput. Phys. Commun.}\ }\textbf {\bibinfo {volume} {183}},\ \bibinfo
  {pages} {324} (\bibinfo {year} {2012})}\BibitemShut {NoStop}%
\bibitem [{SUP()}]{SUPP}%
  \BibitemOpen
  \href@noop {} {}\bibinfo {note} {See Supplemental Material for (i)
  resistivity measurements using the Montgomery method, (ii) quantum
  oscillation data in another sample, and (iii) details of bandstructure
  calculations, which contains~\cite{Walmsley2017Transverse, Montgomery1971Montgomery, Dos2011Montgomery, Perdew1996Calculation, Yates2007Calculation}}\BibitemShut {NoStop}%
\bibitem [{\citenamefont {Walmsley}\ and\ \citenamefont
  {Fisher}(2017)}]{Walmsley2017Transverse}%
  \BibitemOpen
  \bibfield  {author} {\bibinfo {author} {\bibfnamefont {P.}~\bibnamefont
  {Walmsley}}\ and\ \bibinfo {author} {\bibfnamefont {I.}~\bibnamefont
  {Fisher}},\ }\bibfield  {title} {\enquote {\bibinfo {title} {{Determination
  of the resistivity anisotropy of orthorhombic materials via transverse
  resistivity measurements}},}\ }\href@noop {} {\bibfield  {journal} {\bibinfo
  {journal} {Rev. Sci. Instrum.}\ }\textbf {\bibinfo {volume} {88}},\ \bibinfo
  {pages} {043901} (\bibinfo {year} {2017})}\BibitemShut {NoStop}%
\bibitem [{\citenamefont {Montgomery}(1971)}]{Montgomery1971Montgomery}%
  \BibitemOpen
  \bibfield  {author} {\bibinfo {author} {\bibfnamefont {H.}~\bibnamefont
  {Montgomery}},\ }\bibfield  {title} {\enquote {\bibinfo {title} {{Method for
  measuring electrical resistivity of anisotropic materials}},}\ }\href@noop {}
  {\bibfield  {journal} {\bibinfo  {journal} {J. Appl. Phys.}\ }\textbf
  {\bibinfo {volume} {42}},\ \bibinfo {pages} {2971} (\bibinfo {year}
  {1971})}\BibitemShut {NoStop}%
\bibitem [{\citenamefont {Dos~Santos}\ \emph {et~al.}(2011)\citenamefont
  {Dos~Santos}, \citenamefont {De~Campos}, \citenamefont {Da~Luz},
  \citenamefont {White}, \citenamefont {Neumeier}, \citenamefont {De~Lima},\
  and\ \citenamefont {Shigue}}]{Dos2011Montgomery}%
  \BibitemOpen
  \bibfield  {author} {\bibinfo {author} {\bibfnamefont {C.}~\bibnamefont
  {Dos~Santos}}, \bibinfo {author} {\bibfnamefont {A.}~\bibnamefont
  {De~Campos}}, \bibinfo {author} {\bibfnamefont {M.}~\bibnamefont {Da~Luz}},
  \bibinfo {author} {\bibfnamefont {B.}~\bibnamefont {White}}, \bibinfo
  {author} {\bibfnamefont {J.}~\bibnamefont {Neumeier}}, \bibinfo {author}
  {\bibfnamefont {B.}~\bibnamefont {De~Lima}}, \ and\ \bibinfo {author}
  {\bibfnamefont {C.}~\bibnamefont {Shigue}},\ }\bibfield  {title} {\enquote
  {\bibinfo {title} {{Procedure for measuring electrical resistivity of
  anisotropic materials: A revision of the Montgomery method}},}\ }\href@noop
  {} {\bibfield  {journal} {\bibinfo  {journal} {J. Appl. Phys.}\ }\textbf
  {\bibinfo {volume} {110}},\ \bibinfo {pages} {083703} (\bibinfo {year}
  {2011})}\BibitemShut {NoStop}%
\bibitem [{\citenamefont {Perdew}\ \emph {et~al.}(1996)\citenamefont {Perdew},
  \citenamefont {Burke},\ and\ \citenamefont
  {Ernzerhof}}]{Perdew1996Calculation}%
  \BibitemOpen
  \bibfield  {author} {\bibinfo {author} {\bibfnamefont {J.~P.}\ \bibnamefont
  {Perdew}}, \bibinfo {author} {\bibfnamefont {K.}~\bibnamefont {Burke}}, \
  and\ \bibinfo {author} {\bibfnamefont {M.}~\bibnamefont {Ernzerhof}},\
  }\bibfield  {title} {\enquote {\bibinfo {title} {{Generalized gradient
  approximation made simple}},}\ }\href@noop {} {\bibfield  {journal} {\bibinfo
   {journal} {Phys. Rev. Lett.}\ }\textbf {\bibinfo {volume} {77}},\ \bibinfo
  {pages} {3865} (\bibinfo {year} {1996})}\BibitemShut {NoStop}%
\bibitem [{\citenamefont {Yates}\ \emph {et~al.}(2007)\citenamefont {Yates},
  \citenamefont {Wang}, \citenamefont {Vanderbilt},\ and\ \citenamefont
  {Souza}}]{Yates2007Calculation}%
  \BibitemOpen
  \bibfield  {author} {\bibinfo {author} {\bibfnamefont {J.~R.}\ \bibnamefont
  {Yates}}, \bibinfo {author} {\bibfnamefont {X.}~\bibnamefont {Wang}},
  \bibinfo {author} {\bibfnamefont {D.}~\bibnamefont {Vanderbilt}}, \ and\
  \bibinfo {author} {\bibfnamefont {I.}~\bibnamefont {Souza}},\ }\bibfield
  {title} {\enquote {\bibinfo {title} {{Spectral and Fermi surface properties
  from Wannier interpolation}},}\ }\href@noop {} {\bibfield  {journal}
  {\bibinfo  {journal} {Phys. Rev. B.}\ }\textbf {\bibinfo {volume} {75}},\
  \bibinfo {pages} {195121} (\bibinfo {year} {2007})}\BibitemShut {NoStop}%
\bibitem [{\citenamefont {Stevels}\ and\ \citenamefont
  {Wiegers}(1971)}]{Stevels1971CuTe}%
  \BibitemOpen
  \bibfield  {author} {\bibinfo {author} {\bibfnamefont {A.}~\bibnamefont
  {Stevels}}\ and\ \bibinfo {author} {\bibfnamefont {G.}~\bibnamefont
  {Wiegers}},\ }\bibfield  {title} {\enquote {\bibinfo {title} {{Phase
  transitions in copper chalcogenides II. The tellurides Cu$_{3-x}$Te$_2$ and
  CuTe}},}\ }\href@noop {} {\bibfield  {journal} {\bibinfo  {journal} {Rec.
  Trav. Chim. Pays-Bas}\ }\textbf {\bibinfo {volume} {90}},\ \bibinfo {pages}
  {352} (\bibinfo {year} {1971})}\BibitemShut {NoStop}%
\bibitem [{\citenamefont {Shoenberg}(1984)}]{Shoenberg1984Book}%
  \BibitemOpen
  \bibfield  {author} {\bibinfo {author} {\bibfnamefont {D.}~\bibnamefont
  {Shoenberg}},\ }\href@noop {} {\emph {\bibinfo {title} {Magnetic oscillations
  in metals}}}\ (\bibinfo  {publisher} {Cambridge university press,
  Cambridge},\ \bibinfo {year} {1984})\BibitemShut {NoStop}%
\end{thebibliography}
\end{document}